\documentclass[
  aps,
  pre,
  reprint,
  superscriptaddress,
  nofootinbib,
  longbibliography,
  showkeys
]{revtex4-2}

\usepackage{amsmath,amssymb,amsfonts}
\usepackage{bm}
\usepackage{graphicx}
\usepackage{dcolumn}
\usepackage{booktabs}
\usepackage{microtype}
\usepackage{xcolor}
\usepackage[T1]{fontenc} 
\usepackage[english]{babel}
\usepackage{hyperref}
\usepackage{cleveref}

\hypersetup{
  colorlinks=true,
  linkcolor=blue!55!black,
  citecolor=blue!55!black,
  urlcolor=blue!55!black
}

\begin{document}


\title{Cultural, Structural, and Mediated Routes to Polarization\\ in a Nonlinear Mean-Field Model}

\author{Jérôme Michaud}
\email{author.one@example.org}
\affiliation{Department of Business and Mathematics, Mälardalen University, Västerås, Sweden}
\affiliation{Centre for Cultural Evolution, Stockholm University, Stockholm, Sweden}

\author{Fredrik Jansson}
\affiliation{Department of Business and Mathematics, Mälardalen University, Västerås, Sweden}
\affiliation{Centre for Cultural Evolution, Stockholm University, Stockholm, Sweden}
\affiliation{Institute for Futures Studies, Stockholm, Sweden}

\date{\today}

\begin{abstract}
Collective polarization can arise through mechanisms acting at different stages of social communication, but similar polarized outcomes need not imply dynamically equivalent processes. We introduce a nonlinear two-group mean-field model that distinguishes three routes to polarization: individual content filtering, assortativity, and receiver-side message tailoring. Sender-side reformulation is included as a complementary mediation process that transforms source signals before they are mixed. These message transformations are motivated by AI-mediated cultural transmission, in which intermediaries can reformulate messages or adapt them to receivers. Using common and difference coordinates, we derive general conditions for instabilities toward consensus and balanced polarization and analyze the resulting stationary branches in the full two-dimensional phase plane. Individual filtering can generate polarization through state-dependent updating, assortativity can preserve group differences in the social field, and receiver tailoring can create a polarizing feedback even under complete mixing. By contrast, sender-side reformulation affects the polarization mode only when assortativity preserves differentiated source signals. Although these mechanisms can produce identical local instability conditions or balanced-branch amplitudes, they remain distinguishable through transverse stability, asymmetric equilibria, multistability, and basin geometry. In particular, balanced polarization may emerge as a saddle before becoming a stable attractor, while asymmetric saddles organize the boundary between consensus and polarization basins. These results show that collective polarization depends not only on feedback strength, but also on where feedback enters the communication process.
\end{abstract}


\maketitle


\section{Introduction}
\label{sec:introduction}

The formation and persistence of collective polarization are longstanding problems in social dynamics. Statistical-physics models address them by asking how microscopic interaction rules generate macroscopic outcomes such as consensus, fragmentation, coexistence, and polarization \cite{Castellano2009,Lorenz2007,Starnini2026}. Different mechanisms can produce superficially similar polarized states. Bounded-confidence dynamics preserve separated clusters by suppressing interactions between sufficiently dissimilar agents \cite{Deffuant2000,HegselmannKrause2002,Lorenz2007}; attraction--repulsion models can actively amplify large disagreements \cite{SabinMillerAbrams2020,Cornacchia2020}; and nonlinear response, heterogeneity, or group-level feedback can generate distinct consensus and polarization transitions \cite{Baron2021,Gaisbauer2020}. Similar stationary configurations therefore need not imply dynamically equivalent processes. Mechanisms may differ in their onset conditions, nonlinear branches, transverse stability, multistability, and sensitivity to initial conditions. This motivates comparing mechanisms acting at different stages of communication within a common framework. Generative AI extends this problem by allowing intermediaries to transform the messages through which social influence operates \cite{hancock2020aimediated,brinkmann2023machine} .

Existing frameworks distinguish assimilative, similarity-biased, and repulsive social influence \cite{flache2017models}. We organize these mechanisms according to where they act in the communication process. First, \emph{assortativity} determines how strongly social influence is weighted toward similar individuals or members of the same group. This can arise through spatial or social interaction structure, selective exposure, or filtering that favors particular sources. These processes can preserve group-specific information that complete mixing would average away. Yet assortativity alone need not increase opinion divergence under a linear averaging rule; polarization may require its interaction with biased assimilation or another nonlinear response \cite{Dandekar2013}. Cultural-dissemination models likewise show how similarity-dependent interaction can preserve macroscopic diversity despite locally convergent influence \cite{Axelrod1997}. Homophilous argument exchange can also generate polarization without negative influence \cite{mas2013differentiation}. Source filtering can produce selective influence even when encounters themselves are non-assorted, allowing opposing packages of initially unrelated beliefs to emerge \cite{jansson2026emergence}. We represent the aggregate effect of structured encounters and source-dependent filtering through a common assortativity parameter.

Second, \emph{cultural and individual processes} determine how received information is evaluated, revised, adopted, retained, or retransmitted. Cultural-evolutionary theory emphasizes that socially acquired information is not copied indiscriminately: transmission may depend on content, source, context, and the receiver's existing state \cite{BoydRicherson1985,HenrichMcElreath2003,Stubbersfield2022}. Selection can operate at several stages, and cultural transmission may involve reconstruction rather than faithful replication \cite{Stubbersfield2022,Berl2021,Morin2016,claidiere2014how}. Models of cultural systems make this dependence explicit by allowing previously acquired traits to affect the acquisition of further traits through their compatibility relationships \cite{jansson2021modelling}. Related argument-based models connect attitude-dependent evaluation of information to moderation or polarization \cite{banisch2023biased}. The present analysis isolates one minimal representative of this broader class, \emph{individual content filtering}, through which the response to a signal depends on the recipient's current state.

Third, \emph{mediation} transforms the communicated message. We include this component to examine how AI changes cultural transmission: an intermediary can modify information between its expression by a sender and its reception by another individual. Messages can be reformulated on behalf of senders or personalized for receivers, changing the content through which social influence operates \cite{hancock2020aimediated,brinkmann2023machine}. Existing models of algorithmic influence have primarily treated platforms as selectors of information or interaction partners. Such filtering can increase fragmentation, but its collective effect depends on network structure and the microscopic update rule \cite{Sirbu2019,Peralta2021,bellina2023effect}. Whereas algorithmic selection changes which messages are encountered, generative mediation can also change their content. Such transformations may alter the feedback between existing beliefs and incoming information, alongside the effects of individual filtering and assortativity. Generative systems reduce the cost of adapting communication and may alter both message salience and the informational signals associated with its production \cite{Gans2024}. Experiments show that large language models can generate personalized persuasive messages at scale, although their effectiveness depends on the context and design of the interaction \cite{Matz2024,Salvi2025}. Conversational search can also encourage information seeking that reinforces existing views \cite{sharma2024generative}. Conversely, AI mediation can help groups find common ground in deliberation \cite{tessler2024ai}. AI-mediated conversations can also reduce issue or affective polarization under receptive or counterarguing strategies \cite{HruschkaAppel2026}. The relevant theoretical question is therefore not whether personalization is uniformly polarizing, but how receiver-conditioned transformation changes the feedback structure of social influence.

We distinguish two mediation operations. \emph{Sender-side reformulation} changes a message before it is combined with messages from other sources. For example, it may make the position expressed more or less extreme. Under complete mixing, both groups receive the same mixture of reformulated messages. Reformulation can change this common input, but cannot by itself make the groups' inputs different. \emph{Receiver-side tailoring}, by contrast, adapts the delivered message to the receiver's existing position. The same mixture of source messages can therefore yield different inputs to different receivers. This distinction separates transformation before source aggregation from transformation after a receiver has been identified.

We develop a nonlinear two-group mean-field model that places these operations in a common framework. Interaction structure and source-dependent filtering are represented by assortativity, cultural processing by individual content filtering, and mediation by sender-side reformulation and receiver-side tailoring. The model thus contains four operations organized into three conceptual components. It asks which components can independently destabilize neutrality, which require coupling to another component, and whether mechanisms with similar local effects remain distinguishable through their nonlinear dynamics.

The model begins from a bounded scalar coordinate representing a dominant direction of disagreement. This provides a low-dimensional description of cultural-system dynamics. In systems models, beliefs and other cultural traits form repertoires whose internal relationships affect subsequent transmission and acquisition \cite{Buskell2019,jansson2021modelling,jansson2026dynamics}. We reduce this description to a repertoire's position along one main axis of disagreement, retaining the feedback between existing beliefs and responses to new information. Relationships among individual traits enter implicitly through the response functions rather than as separately evolving variables. A companion study develops that richer construction. Here we focus on the reduced collective dynamics and introduce the common and difference coordinates
\begin{equation}
  m=\frac{x_++x_-}{2},
  \qquad
  p=\frac{x_+-x_-}{2},
\end{equation}
where $m$ measures collective displacement and $p$ measures separation between two exchangeable groups. This two-dimensional representation distinguishes common ordering from polarization already at the linear level and permits a complete analysis of neutral, consensus, balanced-polarization, and asymmetric-polarization equilibria.

The analysis yields three main results. First, individual content filtering, assortativity, and receiver-side tailoring can each generate a polarizing instability in an appropriate limit. Sender-side reformulation is not an autonomous route under complete mixing, but it can reinforce or suppress the structural route by transforming the source differences preserved by assortativity. Second, the location of feedback is visible in the linearized difference mode: sender-side reformulation contributes through its product with assortativity, whereas receiver-side tailoring contributes directly even when the sender field is completely mixed. Third, local onset does not determine the nonlinear collective regime. A balanced-polarization branch may emerge as a saddle before becoming stable through a transverse bifurcation, and asymmetric equilibria can organize the boundaries between consensus and polarization basins. Mechanisms that share an instability threshold or balanced-branch amplitude can therefore remain distinct in their transverse stability, multistability, and basin geometry.

The remainder of the paper is organized as follows. Section~\ref{sec:model} defines the individual response, sender-side reformulation, assortativity, receiver-side tailoring, and the two-group reduction. Section~\ref{sec:stationary-states} derives the common- and difference-mode stability conditions and classifies the stationary branches. Section~\ref{sec:three-routes} analyzes the three canonical routes to polarization, and Sec.~\ref{sec:interplay} studies their interactions in two-parameter regime diagrams. Section~\ref{sec:discussion} relates the results to established opinion-dynamics mechanisms and discusses the implications of distinguishing assortativity, individual processing, and message transformation.


\section{Model Definition}
\label{sec:model}

We consider a population divided into two exchangeable groups, labeled $+$ and $-$, whose members hold opinions along a bounded continuous axis. The two-group construction is a minimal closure rather than a claim that empirical populations consist of two homogeneous camps. It separates collective displacement along the opinion axis from differentiation between groups, providing two collective modes whose stationary states and stability can be analyzed explicitly \cite{Castellano2009,Baron2021,Gaisbauer2020}.

The model follows the sequence by which social information is processed and transmitted. An individual response law describes cultural processing; a sender-side map reformulates the source signal; assortative mixing determines how much source differentiation survives aggregation; and receiver-side tailoring conditions the delivered message on the receiver state. These four operations represent three conceptual components: cultural processing, population structure, and mediation. Their order is important because transformation before source aggregation and transformation after receiver identification have different effects on the collective polarization mode.

\subsection{Individual state and cultural processing}
\label{subsec:individual-dynamics}

Each individual carries a scalar state
\begin{equation}
  x_i\in[-1,1],
  \label{eq:individual-state}
\end{equation}
where $x_i=0$ is neutral and the endpoints represent maximally opposed positions. The state changes in response to an effective received signal $r_i\in[-1,1]$ according to
\begin{equation}
  \dot{x}_i=F(x_i,r_i;\boldsymbol{\theta}),
  \label{eq:individual-update-general}
\end{equation}
where $F$ is continuously differentiable and $\boldsymbol{\theta}$ denotes its parameters. We assume opinion-reversal symmetry,
\begin{equation}
  F(-x,-r;\boldsymbol{\theta})=-F(x,r;\boldsymbol{\theta}),
  \label{eq:F-odd-symmetry}
\end{equation}
and require the vector field to point inward at $x=\pm1$, making the interval forward invariant.

We use two response laws. Direct bounded adjustment is
\begin{equation}
  F_{\mathrm{dir}}(x,r)=\lambda(r-x),
  \qquad \lambda>0,
  \label{eq:direct-adjustment}
\end{equation}
where $\lambda$ sets the time scale and may be set to one. This law provides the baseline for polarization generated by population structure or mediation. To represent state-dependent individual content filtering, we use
\begin{equation}
  F_{\mathrm{sel}}(x,r)
  =\frac{\gamma}{2}
  \left[
    r+u(x)-x\bigl(1+r u(x)\bigr)
  \right],
  \label{eq:selective-update}
\end{equation}
with
\begin{equation}
  u(x)=\tanh\!\left(\frac{k_A x}{2}\right).
  \label{eq:acceptance-bias}
\end{equation}
Here $\gamma>0$ sets the update rate and $k_A\geq0$ controls content filtering. At $k_A=0$, Eq.~\eqref{eq:selective-update} becomes $F_{\mathrm{sel}}=(\gamma/2)(r-x)$ and is equivalent to direct adjustment after a rescaling of time. For $k_A>0$, $u(x)$ biases updating toward information congruent with the current state, while the remaining factors preserve boundedness and weaken further motion near $|x|=1$.

Equation~\eqref{eq:selective-update} is a minimal cultural-processing mechanism rather than a model of mediation. It compresses the content- and state-dependent adoption of socially transmitted information into a smooth deterministic response \cite{BoydRicherson1985,HenrichMcElreath2003,Stubbersfield2022}. This abstraction allows individual filtering to be compared directly with differentiated source exposure and message transformation.

\subsection{Message construction and transmission}
\label{subsec:message-construction}

\subsubsection{Sender-side reformulation}
\label{subsec:sender-reformulation}

Before aggregation, a source signal may be reformulated through the odd saturating map
\begin{equation}
  \Phi(x)
  =(1-\rho)x+\rho\tanh(\alpha x),
  \qquad 0\leq\rho\leq1,
  \quad \alpha\geq0.
  \label{eq:reformulation-map}
\end{equation}
The parameter $\rho$ controls the weight of nonlinear reformulation and $\alpha$ its gain and saturation. Its local slope is
\begin{equation}
  \Phi'(0)=(1-\rho)+\rho\alpha.
  \label{eq:reformulation-slope}
\end{equation}
Thus $\alpha>1$ amplifies small signals, $\alpha<1$ compresses them, and $\alpha=1$ leaves the local slope unchanged. The saturation of the hyperbolic tangent bounds this transformation away from neutrality.

Reformulation changes the content or intensity associated with a source, but not the receivers to which that source contributes. It can therefore change the common field under complete mixing, but it cannot by itself create different fields for the two groups.

\subsubsection{Population structure and assortativity}
\label{subsec:sender-mixing}

Let $x_+$ and $x_-$ denote the group mean states. After reformulation, the sender-side fields are
\begin{align}
  z_+
  &=\frac{1+\eta}{2}\Phi(x_+)
   +\frac{1-\eta}{2}\Phi(x_-),\\
  z_-
  &=\frac{1-\eta}{2}\Phi(x_+)
   +\frac{1+\eta}{2}\Phi(x_-),
  \label{eq:sender-fields}
\end{align}
where $0\leq\eta\leq1$ is the fraction of group differentiation preserved by mixing. At $\eta=0$, both groups receive the same population-average field; at $\eta=1$, each receives only its own transformed signal. Intermediate values represent assortative exposure and may summarize homophily, network modularity, selective exposure, or platform-mediated source selection.

The distinction between $\Phi$ and $\eta$ is structural. The map $\Phi$ determines how each source signal is transformed, whereas $\eta$ determines whether differences between transformed source signals survive aggregation. Consequently, sender-side reformulation contributes to the difference mode only when $\eta>0$.

\subsubsection{Receiver-side tailoring and bounded response}
\label{subsec:receiver-tailoring-model}

After source aggregation, the field delivered to each group is conditioned on the receiver state:
\begin{align}
  h_+&=(1-\tau)z_+ + \tau x_+,\\
  h_-&=(1-\tau)z_- + \tau x_-,
  \label{eq:receiver-fields}
\end{align}
where $0\leq\tau\leq1$. At $\tau=0$, the channel input is entirely determined by the sender-side field. At $\tau=1$, it is fully conditioned on the receiver's current group state.

The effective received signal is
\begin{equation}
  r_\pm=G(h_\pm),
  \label{eq:bounded-response-general}
\end{equation}
where $G$ is smooth, odd, increasing, and bounded. We use
\begin{equation}
  G(h)=\tanh(\beta h),
  \qquad \beta>0,
  \label{eq:tanh-response}
\end{equation}
so that $r_\pm\in[-1,1]$ and
\begin{equation}
  G'(0)=\beta.
  \label{eq:response-gain}
\end{equation}
The parameters $\tau$ and $\beta$ play distinct roles: $\tau$ controls how strongly the receiver enters message construction, whereas $\beta$ controls the gain of the bounded communication channel.

Receiver-side tailoring differs from assortative source filtering because it changes the message delivered to a specified receiver rather than the sources entering the social field. It creates the direct feedback loop
\begin{equation}
  x_\pm \longrightarrow h_\pm
  \longrightarrow r_\pm
  \longrightarrow \dot{x}_\pm,
  \label{eq:tailoring-feedback-loop}
\end{equation}
which can amplify group differences even when $z_+=z_-$. This feedback is absent from models in which algorithms modify only partner choice or source exposure \cite{Sirbu2019,Peralta2021}.

\subsection{Two-group mean-field system}
\label{subsec:two-group-reduction}

The group-level dynamics are
\begin{align}
  \dot{x}_+&=F(x_+,r_+;\boldsymbol{\theta}),\\
  \dot{x}_-&=F(x_-,r_-;\boldsymbol{\theta}).
  \label{eq:group-dynamics}
\end{align}
We introduce the common and difference coordinates
\begin{equation}
  m=\frac{x_++x_-}{2},
  \qquad
  p=\frac{x_+-x_-}{2},
  \label{eq:order-parameters-model}
\end{equation}
so that
\begin{equation}
  x_+=m+p,
  \qquad
  x_-=m-p.
  \label{eq:group-states-model}
\end{equation}
The variable $m$ measures collective displacement, while $p$ measures separation between the groups. Thus $p=0$ describes coincident group means, and $m=0$ with $p\neq0$ describes equally strong but opposed group states.

In these coordinates, the sender-side fields are
\begin{align}
  z_+(m,p)
  &=\frac{1+\eta}{2}\Phi(m+p)
   +\frac{1-\eta}{2}\Phi(m-p),\\
  z_-(m,p)
  &=\frac{1-\eta}{2}\Phi(m+p)
   +\frac{1+\eta}{2}\Phi(m-p),
  \label{eq:sender-fields-mp}
\end{align}
and the receiver-conditioned fields are
\begin{align}
  h_+(m,p)&=(1-\tau)z_+(m,p)+\tau(m+p),\\
  h_-(m,p)&=(1-\tau)z_-(m,p)+\tau(m-p).
  \label{eq:receiver-fields-mp}
\end{align}
Using $r_\pm=G(h_\pm)$, the closed mean-field system becomes
\begin{align}
  \dot m
  &=\mathcal{M}(m,p)
  =\frac{1}{2}
  \left[
    F\bigl(m+p,r_+;\boldsymbol{\theta}\bigr)
    +F\bigl(m-p,r_-;\boldsymbol{\theta}\bigr)
  \right],\\
  \dot p
  &=\mathcal{P}(m,p)
  =\frac{1}{2}
  \left[
    F\bigl(m+p,r_+;\boldsymbol{\theta}\bigr)
    -F\bigl(m-p,r_-;\boldsymbol{\theta}\bigr)
  \right].
  \label{eq:mean-field-system-explicit}
\end{align}
Equations~\eqref{eq:reformulation-map}--\eqref{eq:mean-field-system-explicit} define the model.

Since $x_\pm\in[-1,1]$, the physical phase space is the diamond
\begin{equation}
  \mathcal{D}
  =\left\{(m,p)\in\mathbb{R}^2:
  |m|+|p|\leq1\right\}.
  \label{eq:physical-domain-model}
\end{equation}
For both response laws, the vector field points inward at $x_\pm=\pm1$, so $\mathcal{D}$ is forward invariant.

\subsection{Symmetry, invariant axes, and canonical limits}
\label{subsec:symmetry-and-limits}
\label{subsec:symmetry}
\label{subsec:canonical-limits}

The oddness of $F$, $G$, and $\Phi$ gives the global reversal symmetry
\begin{equation}
  (m,p)\mapsto(-m,-p),
  \label{eq:global-reversal}
\end{equation}
while exchange of the two groups gives
\begin{equation}
  (m,p)\mapsto(m,-p).
  \label{eq:group-exchange}
\end{equation}
The consensus axis
\begin{equation}
  p=0
  \label{eq:consensus-axis}
\end{equation}
and the balanced-polarization axis
\begin{equation}
  m=0
  \label{eq:polarization-axis}
\end{equation}
are therefore invariant. The full system may also possess asymmetric polarized states satisfying
\begin{equation}
  m\neq0,
  \qquad
  p\neq0.
  \label{eq:asymmetric-states-definition}
\end{equation}
A state on an invariant axis can be stable along that axis but unstable to transverse perturbations. All equilibria are therefore classified using the full two-dimensional Jacobian.

The three canonical routes studied below are parameter restrictions of the same system.

\paragraph*{Cultural route: individual content filtering.}
We set $\rho=0$, $\eta=0$, and $\tau=0$, and use $F=F_{\mathrm{sel}}$. Both groups receive the same unconditioned social signal, so any group differentiation originates in state-dependent individual updating through $k_A$.

\paragraph*{Structural route: assortative source filtering.}
We set $\tau=0$ and use $F=F_{\mathrm{dir}}$. Assortativity $\eta$ preserves group-specific source information, while $\rho$ and $\alpha$ control how reformulation amplifies or compresses those differences before mixing. Reformulation affects the difference mode only when $\eta>0$.

\paragraph*{Mediation route: receiver-side tailoring.}
We set $\rho=0$ and $\eta=0$, and use $F=F_{\mathrm{dir}}$. The sender-side field is common to both groups, but $\tau>0$ conditions the delivered message on each receiver group. Polarization in this limit is therefore generated by receiver-conditioned transformation rather than differentiated source exposure.

These limits isolate where polarizing feedback enters the communication process while retaining common order parameters and stability criteria. Section~\ref{sec:stationary-states} derives the general stationary-state framework, after which Sec.~\ref{sec:three-routes} compares the three routes beyond their local instability thresholds.


\section{Stationary States and Linear Stability}
\label{sec:stationary-states}

We now derive the stationary-state and stability results needed to compare the cultural, structural, and mediation-based routes to polarization. The analysis is written in terms of the derivatives of the individual response function $F$, the sender-side reformulation map $\Phi$, and the bounded communication map $G$, and therefore applies to both update laws introduced above. The main distinction is between perturbations that displace the two groups together and perturbations that separate them. This mode decomposition identifies the onset of consensus and polarization, while the full two-dimensional Jacobian determines whether the resulting branches are attractors or saddles. Explicit differentiation of the vector field and mechanism-specific expressions are given in Appendices~\ref{app:jacobian}--\ref{app:update-laws}.

\subsection{Stationary states and symmetry classes}
\label{subsec:state-classification}

A stationary state $(m^*,p^*)$ satisfies
\begin{equation}
  \mathcal{M}(m^*,p^*)=0,
  \qquad
  \mathcal{P}(m^*,p^*)=0,
  \label{eq:stationary-mp}
\end{equation}
which is equivalent to
\begin{equation}
  F(x_+^*,r_+^*;\boldsymbol{\theta})=0,
  \qquad
  F(x_-^*,r_-^*;\boldsymbol{\theta})=0,
  \label{eq:stationary-group}
\end{equation}
with $x_+^*=m^*+p^*$ and $x_-^*=m^*-p^*$. The symmetries identified in Sec.~\ref{subsec:symmetry} organize the equilibria into four classes: the neutral state $(0,0)$; consensus states $(m_c,0)$ with $m_c\neq0$; balanced-polarization states $(0,p_b)$ with $p_b\neq0$; and asymmetric-polarization states $(m_a,p_a)$ with $m_a p_a\neq0$. The corresponding labels are
\begin{align}
  (m^*,p^*)&=(0,0),
  \label{eq:neutral-state}\\
  (m^*,p^*)&=(m_c,0), \qquad m_c\neq0,
  \label{eq:consensus-state}\\
  (m^*,p^*)&=(0,p_b), \qquad p_b\neq0,
  \label{eq:balanced-state}\\
  m_a p_a&\neq0.
  \label{eq:asymmetric-state}
\end{align}
Consensus states displace both groups in the same direction, whereas balanced states place them at equal and opposite positions. Asymmetric states combine a nonzero population mean with a nonzero group difference. Group exchange and global reversal generate the symmetry-related set
\begin{equation}
  (m_a,p_a),\quad
  (m_a,-p_a),\quad
  (-m_a,p_a),\quad
  (-m_a,-p_a),
  \label{eq:asymmetric-orbit}
\end{equation}
provided none of these points coincide. Such states are not visible in calculations restricted to either invariant axis.

The stability of any equilibrium is determined from the full Jacobian
\begin{equation}
  J(m,p)
  =\frac{\partial(\mathcal{M},\mathcal{P})}
         {\partial(m,p)}.
  \label{eq:general-jacobian}
\end{equation}
Local asymptotic stability requires
\begin{equation}
  \operatorname{tr}J<0,
  \qquad
  \det J>0,
  \label{eq:trace-determinant-stable}
\end{equation}
whereas $\det J<0$ identifies a saddle. The explicit Jacobian is given in Appendix~\ref{app:jacobian}.

\subsection{Neutral stability and the two collective modes}
\label{subsec:neutral-stability}

At neutrality, symmetry diagonalizes the Jacobian into a common mode $m$, which moves both groups together, and a difference mode $p$, which separates them. Define
\begin{equation}
\begin{split}
    A=F_x(0,0;\boldsymbol{\theta}),
  \qquad
  B=F_r(0,0;\boldsymbol{\theta}),
  \\
  a=\Phi'(0),
  \qquad
  \chi=G'(0).
\end{split}
  \label{eq:neutral-local-derivatives}
\end{equation}
The effective gains of the common and difference fields are
\begin{equation}
  q_{\mathrm{com}}=\tau+(1-\tau)a,
  \qquad
  q_{\mathrm{diff}}=\tau+(1-\tau)a\eta,
  \label{eq:neutral-field-gains}
\end{equation}
and the linearized dynamics are
\begin{align}
  \dot m&=g_{\mathrm{com}}m
  +\mathcal{O}\bigl(\|(m,p)\|^2\bigr),\\
  \dot p&=g_{\mathrm{diff}}p
  +\mathcal{O}\bigl(\|(m,p)\|^2\bigr),
  \label{eq:neutral-linearization}
\end{align}
with
\begin{align}
  g_{\mathrm{com}}
  &=A+B\chi\left[\tau+(1-\tau)a\right],\\
  g_{\mathrm{diff}}
  &=A+B\chi\left[\tau+(1-\tau)a\eta\right].
  \label{eq:general-growth-rates}
\end{align}
The neutral state is locally asymptotically stable if and only if
\begin{equation}
  g_{\mathrm{com}}<0,
  \qquad
  g_{\mathrm{diff}}<0.
  \label{eq:neutral-stability-condition}
\end{equation}

Equation~\eqref{eq:general-growth-rates} is the central local result. Cultural processing enters through the response derivatives $A$ and $B$. Assortativity enters only the difference mode, through $\eta$. Receiver-side tailoring contributes directly through $\tau$, including under complete mixing. By contrast, the reformulation gain $a$ contributes to the difference mode only through the product $a\eta$. Sender-side reformulation can therefore alter common ordering when $\eta=0$, but it cannot by itself destabilize the polarization mode. Receiver-side tailoring can do so because it conditions the received message directly on the receiver state.

The conditions
\begin{equation}
  g_{\mathrm{com}}=0
  \label{eq:common-neutral-boundary}
\end{equation}
and
\begin{equation}
  g_{\mathrm{diff}}=0
  \label{eq:difference-neutral-boundary}
\end{equation}
mark losses of neutral stability in the common and difference directions. Subject to the usual nondegeneracy conditions, they organize pitchfork bifurcations of consensus and balanced-polarization branches, respectively. The linear thresholds locate branch onset but do not determine the criticality of the bifurcation or the stability of the emerging branch.

For direct adjustment, $F(x,r)=r-x$, and $G(h)=\tanh(\beta h)$, the difference-mode threshold reduces to
\begin{equation}
  \beta\left[\tau+(1-\tau)\eta\Phi'(0)\right]=1.
  \label{eq:direct-polarization-threshold}
\end{equation}
This expression makes the distinction between the two mediation operations explicit: sender-side reformulation is filtered by assortativity, whereas receiver-side tailoring remains active at $\eta=0$. For the content-filtered update law in Eq.~\eqref{eq:selective-update},
\begin{equation}
  g_{\mathrm{diff}}
  =\frac{\gamma}{2}
  \left\{
    \frac{k_A}{2}-1
    +\chi\left[\tau+(1-\tau)a\eta\right]
  \right\}.
  \label{eq:selective-difference-rate}
\end{equation}
Here $k_A$ acts through individual processing, $\eta$ controls the preservation of source differences, and $\tau$ supplies direct receiver-conditioned feedback. The corresponding common-mode expression and the required derivatives are collected in Appendix~\ref{app:update-laws}.

\subsection{Invariant branches and transverse stability}
\label{subsec:invariant-branches}

The invariant axes reduce the stationary equations to one dimension, but stability must still be evaluated in the full phase plane. On the consensus axis $p=0$, define
\begin{equation}
  h_c(m)=(1-\tau)\Phi(m)+\tau m,
  \qquad
  r_c(m)=G\bigl(h_c(m)\bigr).
  \label{eq:consensus-fields}
\end{equation}
Consensus states satisfy
\begin{equation}
  F\bigl(m_c,r_c(m_c);\boldsymbol{\theta}\bigr)=0.
  \label{eq:consensus-stationary-equation}
\end{equation}
With
\begin{equation}
\begin{split}
  F_x^c&=F_x\bigl(m_c,r_c(m_c);\boldsymbol{\theta}\bigr),
  \qquad
  F_r^c=F_r\bigl(m_c,r_c(m_c);\boldsymbol{\theta}\bigr),\\
  \phi_c'&=\Phi'(m_c),
  \qquad
  \chi_c=G'\bigl(h_c(m_c)\bigr),
\end{split}
  \label{eq:consensus-local-derivatives}
\end{equation}
the eigenvalues tangent and transverse to the consensus axis are
\begin{align}
  \Lambda_{c,\mathrm{com}}
  &=F_x^c+F_r^c\chi_c
    \left[\tau+(1-\tau)\phi_c'\right],\\
  \Lambda_{c,\mathrm{diff}}
  &=F_x^c+F_r^c\chi_c
    \left[\tau+(1-\tau)\eta\phi_c'\right].
  \label{eq:consensus-eigenvalues}
\end{align}
A consensus state is stable only when both eigenvalues are negative. In particular, $\Lambda_{c,\mathrm{diff}}=0$ marks a transverse loss of stability toward group differentiation and may generate asymmetric polarized states.

On the balanced-polarization axis $m=0$, define
\begin{equation}
  h_b(p)=(1-\tau)\eta\Phi(p)+\tau p,
  \qquad
  r_b(p)=G\bigl(h_b(p)\bigr).
  \label{eq:balanced-fields}
\end{equation}
Balanced states satisfy
\begin{equation}
  F\bigl(p_b,r_b(p_b);\boldsymbol{\theta}\bigr)=0.
  \label{eq:balanced-stationary-equation}
\end{equation}
With
\begin{equation}
\begin{split}
  F_x^b&=F_x\bigl(p_b,r_b(p_b);\boldsymbol{\theta}\bigr),
  \qquad
  F_r^b=F_r\bigl(p_b,r_b(p_b);\boldsymbol{\theta}\bigr),\\
  \phi_b'&=\Phi'(p_b),
  \qquad
  \chi_b=G'\bigl(h_b(p_b)\bigr),
\end{split}
  \label{eq:balanced-local-derivatives}
\end{equation}
the transverse common-mode and tangential difference-mode eigenvalues are
\begin{align}
  \Lambda_{b,\mathrm{com}}
  &=F_x^b+F_r^b\chi_b
    \left[\tau+(1-\tau)\phi_b'\right],\\
  \Lambda_{b,\mathrm{diff}}
  &=F_x^b+F_r^b\chi_b
    \left[\tau+(1-\tau)\eta\phi_b'\right].
  \label{eq:balanced-eigenvalues}
\end{align}
The branch is stable within the polarization axis when $\Lambda_{b,\mathrm{diff}}<0$, but it is an attractor of the full system only when
\begin{equation}
  \Lambda_{b,\mathrm{com}}<0,
  \qquad
  \Lambda_{b,\mathrm{diff}}<0.
  \label{eq:balanced-full-stability}
\end{equation}
Thus a balanced branch may exist and be stable against perturbations along $m=0$ while remaining a saddle in the full phase plane. The transverse condition
\begin{equation}
  \Lambda_{b,\mathrm{com}}=0
  \label{eq:balanced-transverse-boundary}
\end{equation}
marks a stabilization or destabilization of the balanced branch and generically coincides with the creation or absorption of symmetry-related asymmetric branches. These branches may themselves remain saddles, yet their invariant manifolds can organize the boundary between consensus and polarization basins. Derivations of the invariant-axis eigenvalues are provided in Appendix~\ref{app:axis-eigenvalues}.

\subsection{Asymmetric equilibria and dynamical boundaries}
\label{subsec:asymmetric-continuation}

Asymmetric equilibria have no general scalar reduction and must be obtained from
\begin{equation}
  \mathcal{M}(m_a,p_a)=0,
  \qquad
  \mathcal{P}(m_a,p_a)=0,
  \qquad
  m_a p_a\neq0.
  \label{eq:asymmetric-coupled-equations}
\end{equation}
In the numerical analysis, all equilibrium families are located throughout the physical diamond, continued as parameters vary, and classified using the full Jacobian. This procedure captures branches that would be missed by restricting the calculation to $p=0$ or $m=0$; numerical details are given in Appendix~\ref{app:numerical-continuation}.

Three boundaries must be distinguished in the results below. The condition $g_{\mathrm{diff}}=0$ marks the local loss of neutral stability and, generically, the onset of a balanced-polarization branch. The condition $\Lambda_{b,\mathrm{com}}=0$ marks a transverse change in the full stability of an already existing balanced branch. Finally, $\det J=0$ at an asymmetric equilibrium identifies a zero-eigenvalue event, such as a fold or a further symmetry-breaking bifurcation. Keeping these boundaries separate is essential: branch existence, stability within an invariant subspace, and stability in the full phase plane are distinct properties. Consequently, mechanisms that share the same local polarization threshold can still differ in transverse stability, asymmetric equilibria, multistability, and basin geometry.


\section{Routes to Polarization and Their Interactions}
\label{sec:three-routes}
\label{sec:interplay}

We now use the framework of Sec.~\ref{sec:stationary-states} to compare where polarizing feedback enters the communication process and how that location remains visible in the nonlinear dynamics. The cultural route modifies the individual response, the structural route preserves differences between sender fields, and the mediation route conditions the delivered message on the receiver state. Our central comparison separates three properties that need not coincide: the emergence of a nonzero balanced branch, its stability within the invariant axis $m=0$, and its stability against transverse common-mode perturbations. Mechanisms can agree in the first two respects while producing different asymmetric equilibria, multistability, and basin geometry.

All equilibria are classified using the full two-dimensional Jacobian, including asymmetric states away from the invariant axes. In the continuation diagrams, color identifies the equilibrium family, while solid, dashed, and dotted curves denote stable, saddle, and unstable branches, respectively. The general branch equations and stability criteria were derived in Sec.~\ref{sec:stationary-states}; below we retain only the mechanism-specific relations needed to interpret the results.

\subsection{Canonical routes}
\label{subsec:canonical-routes}

\subsubsection{Cultural route: individual content filtering}
\label{subsec:individual-filtering}

To isolate cultural processing, we set $\rho=\eta=\tau=0$ and use the content-filtered response law in Eq.~\eqref{eq:selective-update}. Both groups then receive the same unconditioned social signal, so any differentiation originates in state-dependent individual updating. On the balanced axis, the stationary equation is
\begin{equation}
  p_b=\tanh\!\left(\frac{k_Ap_b}{2}\right),
  \label{eq:filtering-balanced-equation}
\end{equation}
which produces a supercritical balanced branch at
\begin{equation}
  k_A^{\mathrm{em}}=2.
  \label{eq:filtering-emergence}
\end{equation}
The branch is stable within $m=0$ after emergence, but the full stability criterion from Eq.~\eqref{eq:balanced-full-stability} is not satisfied immediately. The communication response still amplifies common-mode perturbations, so the new balanced states initially remain saddles in the full phase plane.

\begin{figure*}[t!]
  \centering
  \includegraphics[width=\textwidth]{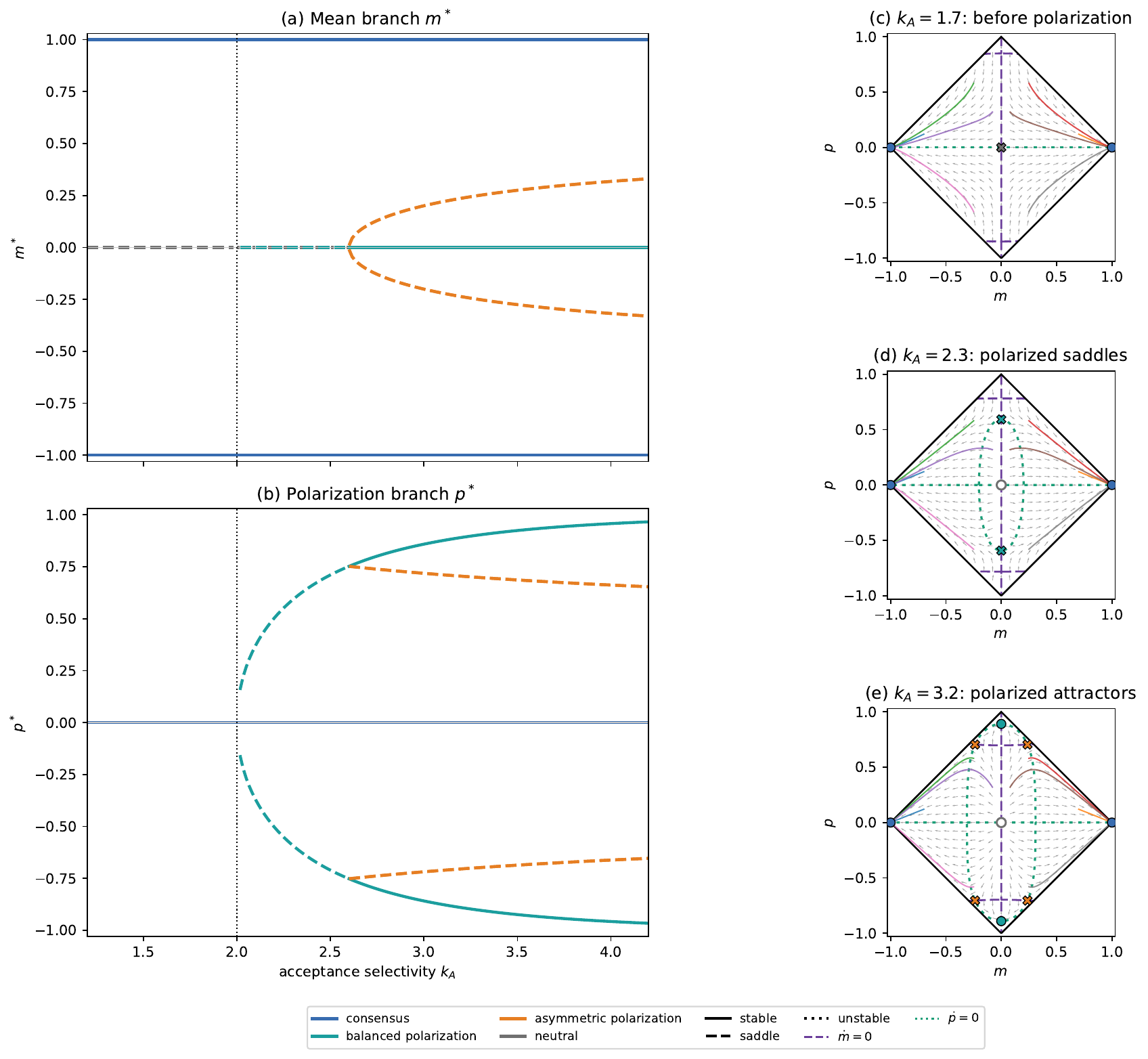}
  \caption{Cultural route to polarization through individual content filtering. The left column shows the complete continuation of $m^*$ and $p^*$ as $k_A$ varies. The right column shows representative phase portraits before the balanced branch emerges, while it is stable only within the polarization axis, and after transverse stabilization. Colors distinguish neutral, consensus, balanced-polarization, and asymmetric-polarization equilibria. Solid, dashed, and dotted curves denote stable, saddle, and unstable branches, respectively, as determined from the full two-dimensional Jacobian.}
  \label{fig:individual-filtering-results}
\end{figure*}

Figure~\ref{fig:individual-filtering-results} shows the resulting two-stage sequence. Increasing $k_A$ first creates the balanced branch and subsequently moves it into a regime where saturation weakens the local response enough for transverse stabilization. Asymmetric saddles occur near this secondary transition and help separate the consensus and balanced-polarization basins. Individual filtering is therefore an autonomous route to a polarized branch, but the existence of that branch does not by itself imply stable collective polarization.

\subsubsection{Structural route: assortative source filtering}
\label{subsec:assortative-reformulation}

For the structural route, we set $\tau=0$ and use direct adjustment. Assortativity $\eta$ preserves group-specific source information, while the reformulation map in Eq.~\eqref{eq:reformulation-map} modifies that information before mixing. The neutral difference mode loses stability when
\begin{equation}
  \beta\eta\left[(1-\rho)+\rho\alpha\right]=1.
  \label{eq:reformulation-emergence}
\end{equation}
This product is the defining feature of the structural route. Reformulation changes the gain of a source difference, but assortativity determines whether that difference survives aggregation. Increasing $\alpha$ can therefore lower the assortativity required for polarization, whereas no finite reformulation gain creates a difference mode at $\eta=0$.

\begin{figure*}[t!]
  \centering
  \includegraphics[width=\textwidth]{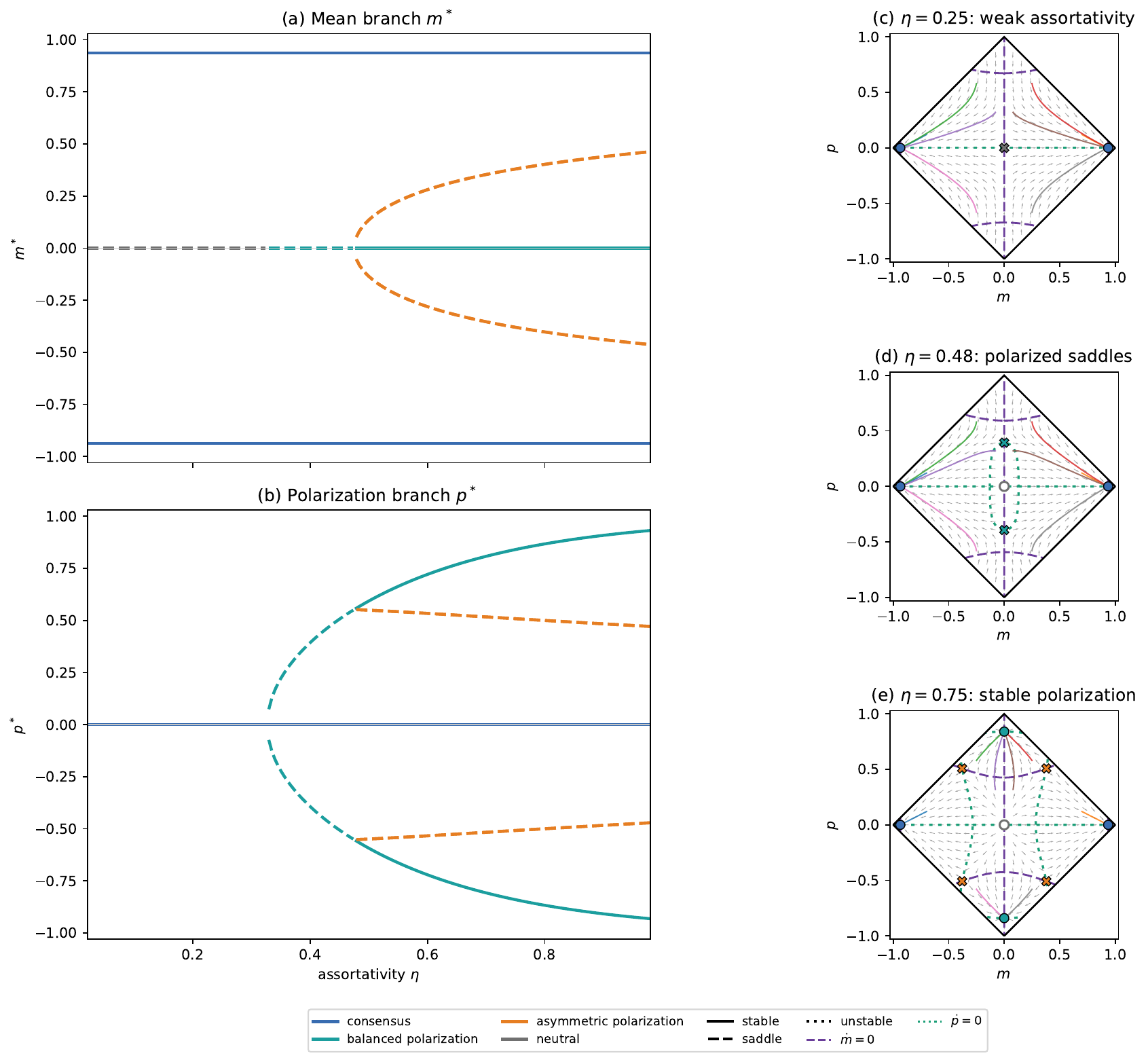}
  \caption{Structural route to polarization through assortative source filtering, modified by sender-side reformulation. Complete equilibrium branches are continued in $\eta$ for fixed reformulation weight, reformulation nonlinearity, and communication gain. The phase portraits illustrate the progression from no nonzero balanced branch, through balanced polarized saddles, to stable balanced polarization. Sender-side reformulation transforms the source signals, whereas $\eta$ determines whether their difference survives mixing.}
  \label{fig:reformulation-results}
\end{figure*}

Figure~\ref{fig:reformulation-results} shows that crossing the emergence boundary first creates balanced states that remain saddles in the full phase plane. As $\eta$ increases, their amplitude grows and nonlinear saturation eventually stabilizes the transverse direction. Asymmetric equilibria again occur near this change of stability and contribute to the basin boundary. Sender-side reformulation is therefore not an autonomous fourth route: it modifies the strength and saturation of the structural route, which still depends on preserving source differences.

\subsubsection{Mediation route: receiver-side tailoring}
\label{subsec:receiver-tailoring-results}

\begin{figure*}[t!]
  \centering
  \includegraphics[width=\textwidth]{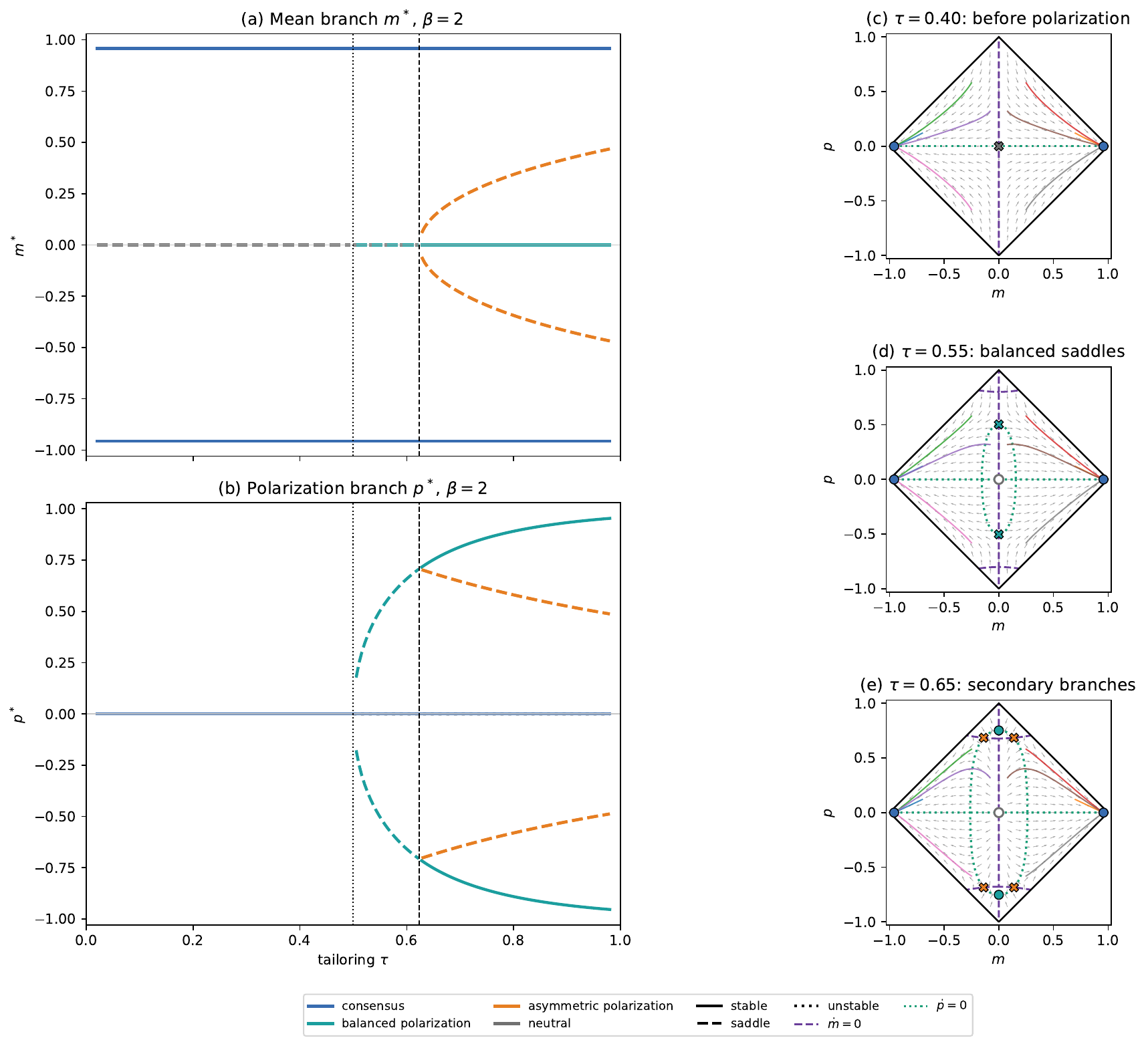}
  \caption{Mediation route to polarization through receiver-side tailoring for $\beta=2$. The left column gives the complete continuation of $m^*$ and $p^*$ in $\tau$. The right column shows phase portraits before branch emergence, while the balanced branch is a saddle, and after its transverse stabilization. The balanced branch appears at $\tau=1/2$ and becomes stable at $\tau\simeq0.6232$. The sender-side field is common to both groups throughout.}
  \label{fig:tailoring-results}
\end{figure*}

To isolate receiver-conditioned mediation, we use direct adjustment and set $\rho=\eta=0$. The sender-side field is common to both groups, but tailoring introduces a group-specific contribution after aggregation. The balanced branch satisfies
\begin{equation}
  p_b=\tanh(\beta\tau p_b)
  \label{eq:tailoring-balanced-equation-results}
\end{equation}
and emerges at
\begin{equation}
  \beta\tau=1.
  \label{eq:tailoring-emergence-results}
\end{equation}
Using the transverse condition from Sec.~\ref{subsec:invariant-branches}, its full-stability boundary is
\begin{equation}
  \tau_{\mathrm{stab}}
  =\frac{\operatorname{arctanh}\sqrt{1-1/\beta}}
  {\beta\sqrt{1-1/\beta}}.
  \label{eq:tailoring-stability-threshold-results}
\end{equation}
For the value $\beta=2$ used in Fig.~\ref{fig:tailoring-results}, branch emergence occurs at $\tau=1/2$ and transverse stabilization at $\tau\simeq0.6232$.

The continuation again displays a balanced branch that appears before it becomes a full attractor. Beyond the secondary transition, stable consensus and stable balanced polarization coexist, with asymmetric saddles contributing to their basin boundary. Unlike sender-side reformulation, however, receiver tailoring operates under complete mixing. Because it is applied after source aggregation and conditioned on the receiver, it creates a difference mode even though both groups begin from the same sender-side field.

\subsubsection{What the canonical routes share and what distinguishes them}
\label{subsec:canonical-comparison}

The three routes exhibit the same qualitative two-stage sequence in the parameter regimes considered here: a balanced branch emerges before it becomes stable against transverse perturbations. This shared sequence does not make the mechanisms dynamically equivalent. Individual filtering changes the response derivatives, assortativity controls the survival of source differences, and tailoring creates receiver-conditioned feedback after aggregation. These locations determine which parameters act directly on the difference mode and how the common mode behaves along the polarized branch.

The canonical continuations therefore establish two complementary results. First, cultural filtering, structural differentiation, and receiver tailoring can each support polarization in an appropriate limit, while sender-side reformulation requires assortativity. Second, a local difference-mode instability is not sufficient to predict the collective regime. The transverse eigenvalue and the associated asymmetric equilibria determine whether balanced polarization is a saddle or an attractor and how its basin is embedded among competing consensus states.

\subsection{Interactions and partial dynamical equivalence}
\label{subsec:route-interactions}

The pairwise interaction diagrams test how far one mechanism can substitute for another. Dashed curves mark balanced-branch emergence and solid curves transverse stabilization. The distinction between these curves is central: mechanisms can be interchangeable in the stationary equation while remaining distinguishable in the full phase plane.

\subsubsection{Individual filtering and receiver-side tailoring}
\label{subsec:filtering-tailoring}

With $\eta=\rho=0$, the balanced stationary equation depends only on the combined gain
\begin{equation}
  \frac{k_A}{2}+\beta\tau,
  \qquad
  \frac{k_A}{2}+\beta\tau=1
  \quad\text{at emergence}.
  \label{eq:filtering-tailoring-emergence}
\end{equation}
Parameter combinations with the same combined gain consequently generate the same balanced-branch amplitude. Cultural filtering and receiver tailoring are therefore interchangeable at the level of branch existence.

\begin{figure*}[t!]
  \centering
  \includegraphics[width=\textwidth]{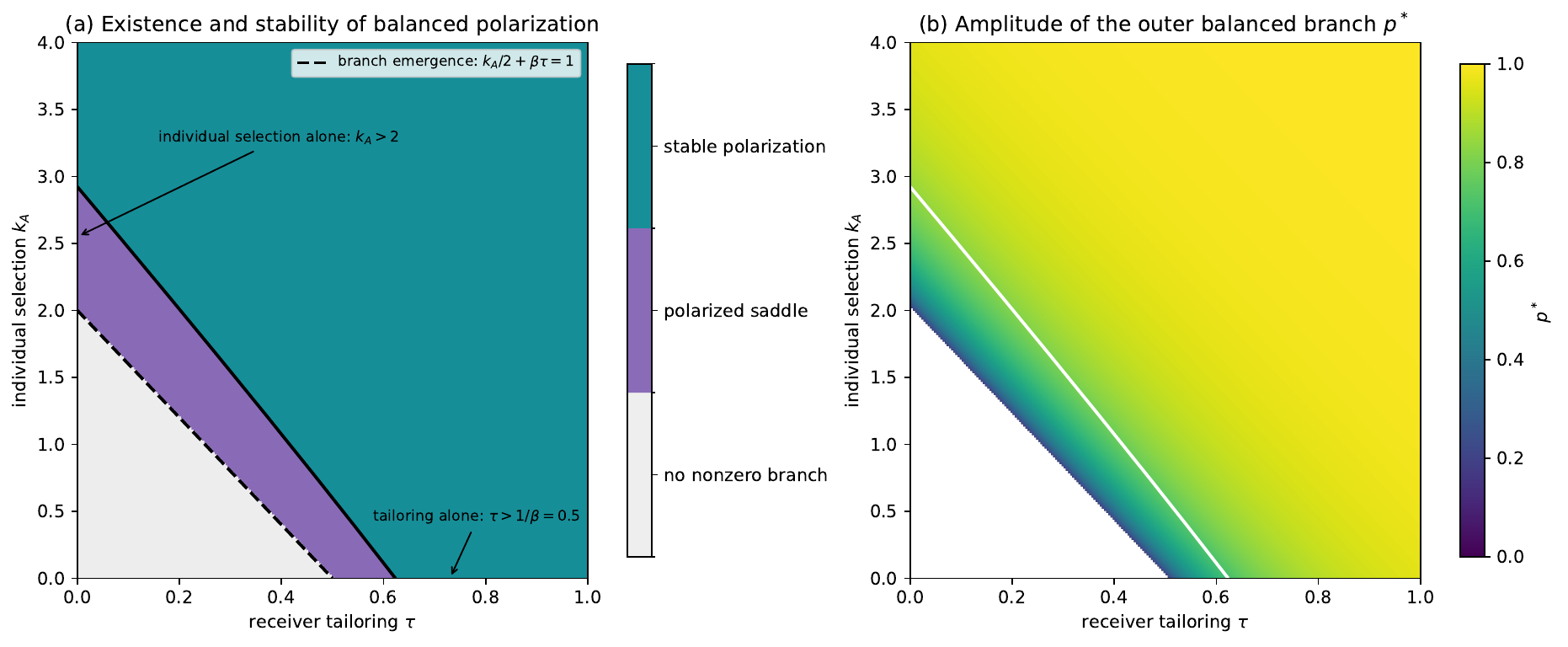}
  \caption{Interaction between individual content filtering and receiver-side tailoring for $\beta=2$. The dashed line is the exact emergence condition $k_A/2+\beta\tau=1$, and the solid line is the transverse-stability boundary. Filtering and tailoring are interchangeable with respect to the balanced branch equation and its amplitude, but not with respect to stability against common-mode perturbations.}
  \label{fig:filtering-tailoring-plane}
\end{figure*}

The equivalence ends at transverse stability. The parameter $k_A$ modifies the individual response law, whereas $\tau$ modifies the receiver-conditioned field. Their contributions to the full Jacobian differ even at matched $p_b$. Figure~\ref{fig:filtering-tailoring-plane} shows this directly: the emergence line follows the additive effective gain, but the stability boundary retains information about where the feedback enters.

\subsubsection{Assortativity and receiver-side tailoring}
\label{subsec:assortativity-tailoring}

For direct adjustment and a linear sender map, assortativity and tailoring enter the balanced branch through the effective difference gain
\begin{equation}
  q(\tau,\eta)=\tau+(1-\tau)\eta.
  \label{eq:q-tau-eta-results}
\end{equation}
Branch emergence is given by $\beta q=1$, or equivalently
\begin{equation}
  \eta_{\mathrm{em}}(\tau)
  =\frac{1/\beta-\tau}{1-\tau}.
  \label{eq:tau-eta-emergence-curve}
\end{equation}
Thus assortativity and tailoring can compensate for one another in the difference mode, but not symmetrically. The assortative contribution is weighted by $1-\tau$ because receiver conditioning progressively replaces the sender-side field. At complete receiver conditioning, source composition becomes irrelevant; tailoring, by contrast, can support a difference mode at $\eta=0$.

\begin{figure*}[t!]
  \centering
  \includegraphics[width=\textwidth]{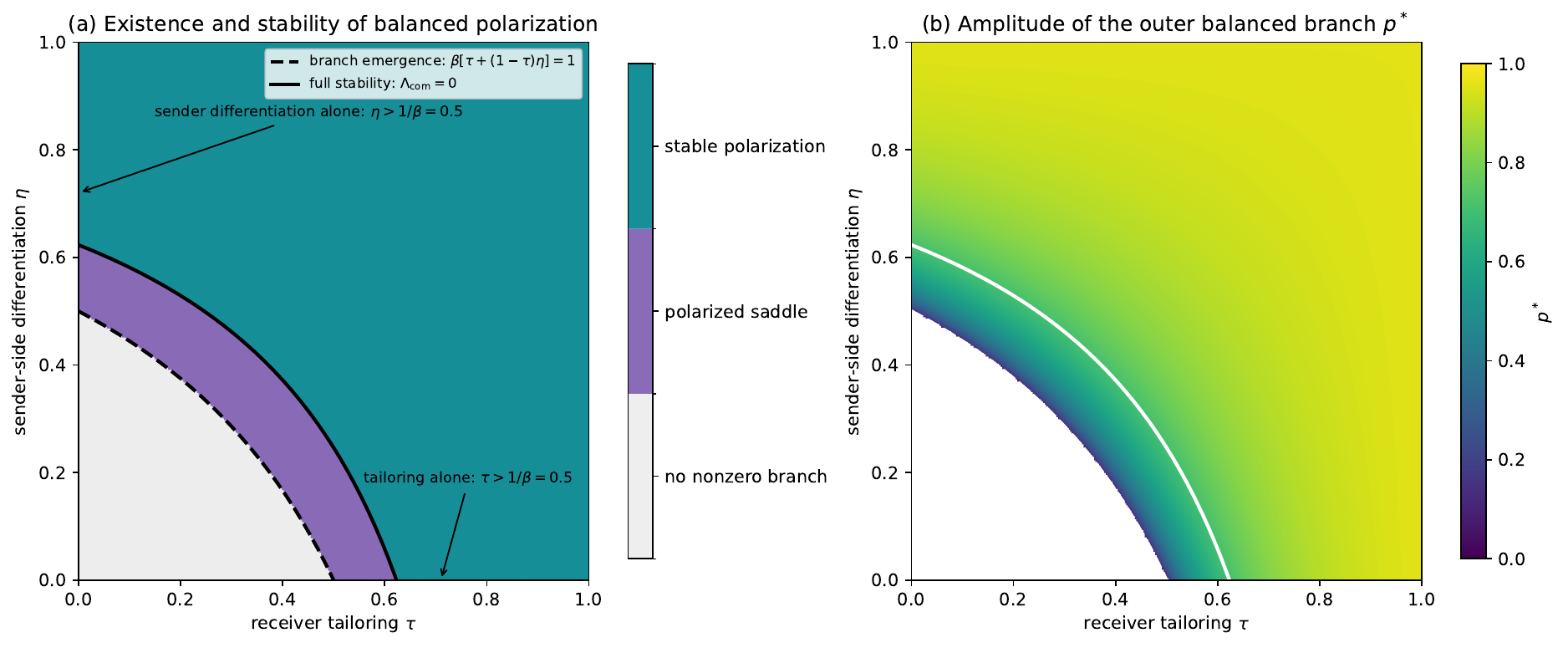}
  \caption{Interaction between receiver-side tailoring and assortative source filtering for a linear sender map and $\beta=2$. The effective difference gain is $q=\tau+(1-\tau)\eta$. The dashed and solid curves are the exact branch-emergence and transverse-stability boundaries, respectively. Assortativity acts through the sender-side field and is progressively displaced as receiver conditioning becomes dominant.}
  \label{fig:tailoring-assortativity-plane}
\end{figure*}

Figure~\ref{fig:tailoring-assortativity-plane} shows that the emergence and transverse-stability boundaries inherit this directional replacement. The mechanisms can match the balanced-state amplitude through $q$, but their interpretation and intervention points remain different.

\subsubsection{Assortativity and sender-side reformulation}
\label{subsec:reformulation-assortativity}

The final interaction occurs entirely before receiver identification. With $\tau=0$ and fixed $\rho$, the emergence boundary is
\begin{equation}
  \eta_{\mathrm{em}}(\alpha)
  =\frac{1}{\beta[(1-\rho)+\rho\alpha]}.
  \label{eq:eta-alpha-emergence-curve}
\end{equation}
Increasing the reformulation gain lowers the assortativity required for branch onset and can indirectly promote transverse stabilization by increasing branch amplitude and saturation. Nevertheless, Eq.~\eqref{eq:eta-alpha-emergence-curve} has no finite solution at $\eta=0$. Reformulation amplifies or compresses source differences but cannot replace the structural process that preserves them.

\begin{figure*}[t!]
  \centering
  \includegraphics[width=\textwidth]{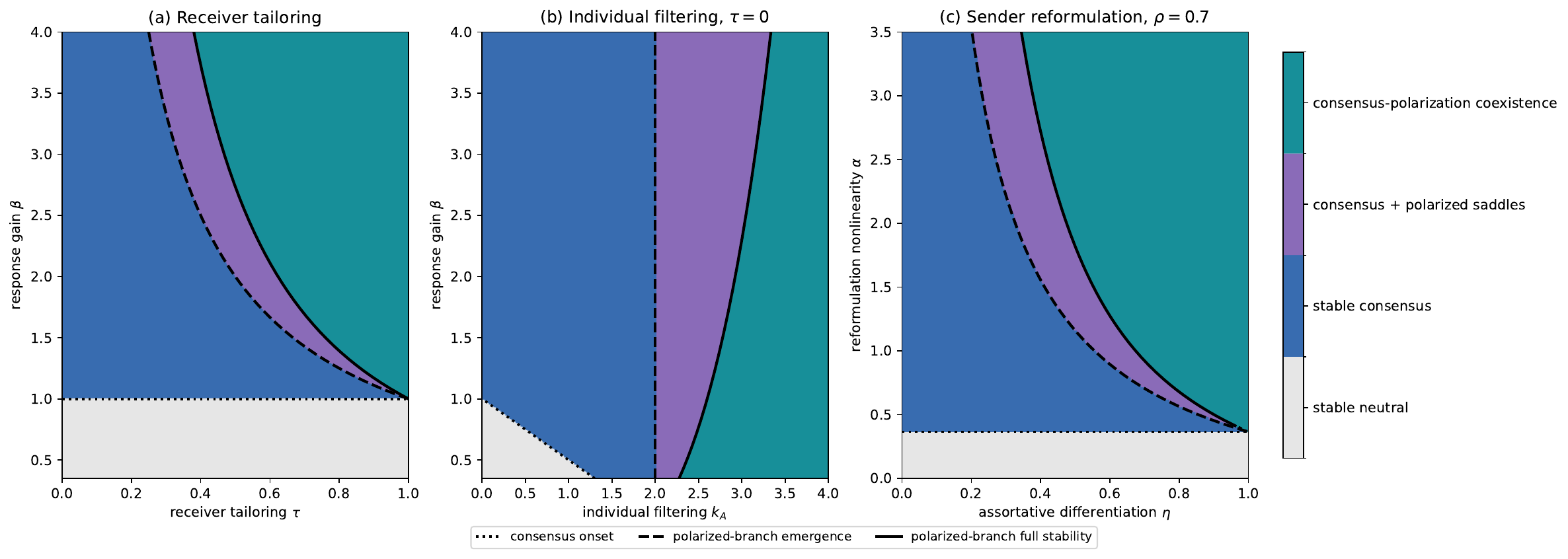}
  \caption{Phase structure for the three canonical routes and the sender-side structural interaction. Panel (a) varies receiver-side tailoring $\tau$ and communication gain $\beta$. Panel (b) varies individual content filtering $k_A$ and $\beta$ with $\tau=0$. Panel (c) varies assortative source filtering $\eta$ and sender-side reformulation gain $\alpha$ with $\tau=0$, $\rho=0.7$, and $\beta=1.8$. Gray denotes stable neutrality, blue stable consensus without a nonzero balanced branch, purple stable consensus coexisting with balanced polarized saddles, and teal coexistence of stable consensus and stable balanced polarization. Dotted curves mark consensus onset, dashed curves balanced-branch emergence, and solid curves full transverse stabilization.}
  \label{fig:three-route-phase-revised}
\end{figure*}

Figure~\ref{fig:three-route-phase-revised} places this sender-side interaction beside the corresponding phase structures for receiver tailoring and individual filtering. The comparison condenses the main result of the paper. Feedback in the individual response and feedback conditioned on the receiver can each generate polarization directly. Reformulation before aggregation affects the polarization mode only when assortativity retains differentiated source signals. In all cases, branch emergence and transverse stabilization remain separate boundaries.

\subsection{Nonlinear synthesis}
\label{subsec:nonlinear-synthesis}

The results identify three levels at which mechanisms can be compared. The linear difference-mode gain determines the local instability of neutrality. The balanced stationary equation determines the existence and amplitude of a polarized branch. The transverse eigenvalue and off-axis equilibria determine whether that branch is a collective attractor and how its basin is organized. Agreement at either of the first two levels does not imply agreement at the third.

This hierarchy explains why the location of feedback remains dynamically observable even when its effective strength is matched. Cultural filtering acts through the individual response, assortativity through the preservation of source differences, reformulation through the transformation of those preserved signals, and tailoring through receiver-conditioned message construction. Their positions in the communication process remain encoded in the full Jacobian. In the regimes examined here, asymmetric saddles provide a nonlinear manifestation of these differences by contributing to the boundary between consensus and balanced-polarization basins. Collective polarization is therefore determined not only by whether a polarized state exists, but also by how it is created, stabilized, and embedded in the global phase portrait.


\section{Discussion}
\label{sec:discussion}

This paper provides a unified dynamical comparison of three routes to collective polarization that act at different stages of communication. Individual content filtering changes how received information is processed; assortativity changes which group-specific signals survive aggregation; and receiver-side tailoring changes the message delivered to a recipient. Sender-side reformulation is a fourth modeled operation, but not an autonomous route under complete mixing: its effect on differences between group inputs depends on assortativity preserving differentiated source signals. This separation between cultural processing, assortative influence, and message transformation is the main conceptual contribution of the model.

The comparison brings together mechanisms studied in social influence \cite{flache2017models} and cultural evolution models \cite{jansson2026dynamics,jansson2026modelling}, examining how they generate collective agreement or differentiation. State-dependent content filtering is related to biased assimilation and to cultural-transmission models in which adoption depends on content, context, and the receiver's existing state \cite{Dandekar2013,BoydRicherson1985,HenrichMcElreath2003,Stubbersfield2022,jansson2021modelling}. Related models explicitly connect internal attitude organization with between-person influence, as in the hierarchical Ising opinion model \cite{vandermaas2020polarization}. Assortativity summarizes the differentiation of social input arising from structured encounters or selective influence, including source filtering under non-assorted encounters \cite{jansson2026emergence}. Such differentiation need not generate increasing extremity under a linear averaging rule, and its collective effects depend on how information is processed \cite{Dandekar2013,Sirbu2019,Peralta2021}. Receiver-side tailoring instead represents transformation conditional on the recipient. This mechanism is increasingly relevant when generative systems can adapt wording, framing, or emphasis rather than merely rank preexisting content \cite{Matz2024,Salvi2025}. The model therefore extends the usual distinction between individual response and interaction structure by separating assortative influence from sender-side and receiver-side message transformation.

The clearest analytical distinction concerns where transformation occurs relative to source aggregation. Sender-side reformulation enters the difference mode through the product $\eta\Phi'(0)$. It can lower the assortativity required for polarization, but under complete mixing, both groups receive the same mixture of reformulated messages. Changing that mixture can affect their common input, but cannot by itself differentiate their inputs. Receiver-side tailoring acts after aggregation and therefore contributes directly to the difference mode even at $\eta=0$. This result does not imply that personalization is intrinsically polarizing. Empirical studies find that personalized generative messages can increase persuasion \cite{Matz2024,Salvi2025}, although personalization does not consistently outperform untailored messages \cite{hackenburg2024evaluating}. Appropriately designed AI mediation can also promote common ground or reduce issue or affective polarization \cite{tessler2024ai,HruschkaAppel2026}. These experiments motivate studying message transformation, but do not establish its long-run population dynamics. The theoretical claim is conditional: receiver-specific transformation creates a feedback channel capable of amplifying differences without assortative exposure, and its realized effect depends on the response law, gain, and communicative objective.

A second contribution is to distinguish the onset of a polarized branch from stable polarization. Across the three routes, a nonzero balanced solution may emerge while remaining unstable to common-mode perturbations. Loss of neutral stability, existence of a balanced branch, and stabilization in the full phase plane are therefore separate events. Mechanisms can even be interchangeable at the level of the balanced stationary equation without being dynamically equivalent. Individual filtering and receiver tailoring, for example, combine additively in the branch-emergence condition, yet they enter different derivatives of the vector field and need not produce the same transverse stability. Matching a stationary amplitude is consequently insufficient to identify the process that generated it.

The full two-dimensional analysis also reveals asymmetric polarized equilibria that are invisible on the consensus and balanced-polarization axes. In the cases considered here, these states commonly arise near transverse bifurcations and act as saddles whose invariant manifolds help separate consensus and polarization basins. Stable consensus and stable balanced polarization can therefore coexist, making the long-run outcome dependent on initial conditions as well as parameter values. This nonlinear organization strengthens the comparison with earlier mean-field and attraction--repulsion models in which several collective states coexist \cite{Baron2021,Gaisbauer2020,SabinMillerAbrams2020}. More generally, it shows why a mechanism should be characterized not only by its local instability threshold, but also by the branch structure and basin geometry it produces.

The model is intentionally minimal. The two-group closure tracks group means rather than within-group distributions and cannot represent multiple clusters, endogenous group formation, or arbitrary network topology. The dynamics are deterministic and symmetric, and the parameters summarize processes that may be difficult to separate empirically. In particular, $\eta$ is an effective measure of retained source differentiation, whether arising from interaction structure or source-dependent filtering. It is held fixed rather than generated through evolving networks or belief-dependent assessments of sources. $\Phi$ compresses heterogeneous sender-side transformations into a scalar map, and $\tau$ measures receiver dependence rather than a directly observable property of a particular platform or language model. The content-filtered update law is likewise a reduced representation of cultural processing, not a complete model of innovation, memory, revision, and retransmission \cite{Stubbersfield2022,Buskell2019,jansson2021modelling,jansson2026dynamics}. These simplifications permit a complete bifurcation analysis, but empirical application will require richer population structure, stochasticity, and measurements that distinguish original messages, transformed outputs, exposure patterns, and subsequent opinion change.

The central conclusion is that different routes to polarization can produce similar outcomes without being dynamically equivalent. Individual filtering changes how incoming information affects an individual, assortativity changes the composition of social input, and mediation transforms the messages being transmitted. Their effects can coincide at the onset of an instability or in the amplitude of a balanced polarized state, while differing in transverse stability, asymmetric equilibria, and basin geometry. Explaining polarization therefore requires identifying both where feedback enters the communication process and whether the resulting polarized states are stable against more general perturbations.

AI-mediated communication provides an important application of this distinction. By reformulating messages or adapting them to receivers, AI can modify familiar feedback processes and introduce additional routes to polarization. The contrast between sender reformulation and receiver tailoring shows why these transformations should be modeled explicitly when assessing how AI changes cultural transmission.


\section*{Data Availability}
No empirical data were used in this study. All numerical results were generated from the mathematical model described in the article. The source code used to generate the numerical results and figures will be made available in a public repository upon publication.


\appendix


\section{Full Jacobian of the Mean-Field System}
\label{app:jacobian}

This appendix gives the explicit Jacobian underlying the stability results in Sec.~\ref{sec:stationary-states}. Define
\begin{equation}
  f_\pm(m,p)
  =F\bigl(x_\pm,r_\pm;\boldsymbol{\theta}\bigr),
  \qquad
  x_\pm=m\pm p.
  \label{eq:app-fpm-definition}
\end{equation}
The mean-field vector field can be written as
\begin{equation}
  \begin{pmatrix}
    \mathcal{M}\\
    \mathcal{P}
  \end{pmatrix}
  =\frac{1}{2}
  \begin{pmatrix}
    f_++f_-\\
    f_+-f_-
  \end{pmatrix},
  \label{eq:app-mean-difference-transform}
\end{equation}
so that
\begin{equation}
  J(m,p)
  =\frac{1}{2}
  \begin{pmatrix}
    \partial_m f_+ + \partial_m f_- &
    \partial_p f_+ + \partial_p f_-\\[2mm]
    \partial_m f_+ - \partial_m f_- &
    \partial_p f_+ - \partial_p f_-
  \end{pmatrix}.
  \label{eq:app-general-jacobian-expanded}
\end{equation}

Let
\begin{equation}
  F_x^\pm
  =\left.\frac{\partial F}{\partial x}\right|_{(x_\pm,r_\pm)},
  \qquad
  F_r^\pm
  =\left.\frac{\partial F}{\partial r}\right|_{(x_\pm,r_\pm)},
  \label{eq:app-F-derivatives}
\end{equation}
with $G_\pm'=G'(h_\pm)$ and $\Phi_\pm'=\Phi'(x_\pm)$. The chain rule gives
\begin{align}
  \partial_m f_+
  &=F_x^+ + F_r^+G_+'\,\partial_m h_+,
  \\
  \partial_p f_+
  &=F_x^+ + F_r^+G_+'\,\partial_p h_+,
  \\
  \partial_m f_-
  &=F_x^- + F_r^-G_-'\,\partial_m h_-,
  \\
  \partial_p f_-
  &=-F_x^- + F_r^-G_-'\,\partial_p h_-.
  \label{eq:app-f-derivatives}
\end{align}
The sender-side field derivatives are
\begin{align}
  \partial_m z_+
  &=\frac{1+\eta}{2}\Phi_+'+
    \frac{1-\eta}{2}\Phi_-',
  \\
  \partial_p z_+
  &=\frac{1+\eta}{2}\Phi_+'-
    \frac{1-\eta}{2}\Phi_-',
  \\
  \partial_m z_-
  &=\frac{1-\eta}{2}\Phi_+'+
    \frac{1+\eta}{2}\Phi_-',
  \\
  \partial_p z_-
  &=\frac{1-\eta}{2}\Phi_+'-
    \frac{1+\eta}{2}\Phi_-'.
  \label{eq:app-z-derivatives}
\end{align}
Since $h_\pm=(1-\tau)z_\pm+\tau x_\pm$, their derivatives are
\begin{align}
  \partial_m h_+
  &=(1-\tau)\partial_m z_+ + \tau,
  \\
  \partial_p h_+
  &=(1-\tau)\partial_p z_+ + \tau,
  \\
  \partial_m h_-
  &=(1-\tau)\partial_m z_- + \tau,
  \\
  \partial_p h_-
  &=(1-\tau)\partial_p z_- - \tau.
  \label{eq:app-h-derivatives}
\end{align}
Equations~\eqref{eq:app-general-jacobian-expanded}--\eqref{eq:app-h-derivatives} give the full Jacobian at arbitrary points in the physical state space.

At neutrality, odd symmetry implies $\Phi_+'=\Phi_-'=a$, $G_+'=G_-'=\chi$, and identical derivatives of $F$ in the two groups. Substitution into Eq.~\eqref{eq:app-general-jacobian-expanded} eliminates the off-diagonal terms and gives the common- and difference-mode growth rates reported in Eq.~\eqref{eq:general-growth-rates}.

\section{Invariant-Axis Eigenvalues}
\label{app:axis-eigenvalues}

\subsection{Consensus branch}

On the consensus axis, $p=0$ and $x_+=x_-=m_c$. The two source states are identical, so
\begin{equation}
  z_+=z_-=\Phi(m_c),
  \qquad
  h_+=h_-=h_c(m_c).
\end{equation}
The common perturbation changes the two source states in the same direction. Its field derivative is
\begin{equation}
  \partial_{
m, \mathrm{com}}h
  =\tau+(1-\tau)\Phi'(m_c),
\end{equation}
whereas a difference perturbation is filtered by assortativity,
\begin{equation}
  \partial_{
m, \mathrm{diff}}h
  =\tau+(1-\tau)\eta\Phi'(m_c).
\end{equation}
Applying the chain rule to the individual response gives
\begin{align}
  \Lambda_{c,\mathrm{com}}
  &=F_x^c+F_r^c\chi_c
  \left[\tau+(1-\tau)\phi_c'\right],
  \\
  \Lambda_{c,\mathrm{diff}}
  &=F_x^c+F_r^c\chi_c
  \left[\tau+(1-\tau)\eta\phi_c'\right],
\end{align}
which reproduces Eq.~\eqref{eq:consensus-eigenvalues}.

\subsection{Balanced-polarization branch}

On the balanced axis, $m=0$, $x_+=p_b$, and $x_-=-p_b$. Oddness of $\Phi$ and $G$ gives
\begin{equation}
  z_+=\eta\Phi(p_b),
  \qquad
  z_-=-\eta\Phi(p_b),
  \qquad
  h_-=-h_+.
\end{equation}
Because the derivatives of odd functions are even, the local derivatives coincide in the two groups. A common perturbation is not attenuated by assortative mixing, whereas a difference perturbation is. The corresponding field derivatives are
\begin{equation}
\begin{split}
    \partial_{
m, \mathrm{com}}h
  &=\tau+(1-\tau)\Phi'(p_b),
  \\
  \partial_{
m, \mathrm{diff}}h
  &=\tau+(1-\tau)\eta\Phi'(p_b).
\end{split}
\end{equation}
The eigenvalues are therefore
\begin{align}
  \Lambda_{b,\mathrm{com}}
  &=F_x^b+F_r^b\chi_b
  \left[\tau+(1-\tau)\phi_b'\right],
  \\
  \Lambda_{b,\mathrm{diff}}
  &=F_x^b+F_r^b\chi_b
  \left[\tau+(1-\tau)\eta\phi_b'\right],
\end{align}
which reproduces Eq.~\eqref{eq:balanced-eigenvalues}.

\section{Specialization to the Update Laws}
\label{app:update-laws}

\subsection{Direct adjustment}

For direct adjustment with the time scale set to one,
\begin{equation}
  F(x,r)=r-x,
\end{equation}
so that
\begin{equation}
  F_x=-1,
  \qquad
  F_r=1.
  \label{eq:app-direct-F-derivatives}
\end{equation}
The neutral growth rates are
\begin{align}
  g_{\mathrm{com}}
  &=-1+\chi
  \left[\tau+(1-\tau)a\right],
  \\
  g_{\mathrm{diff}}
  &=-1+\chi
  \left[\tau+(1-\tau)a\eta\right].
  \label{eq:app-direct-neutral-rates}
\end{align}
For $G(h)=\tanh(\beta h)$, $\chi=\beta$, and $a=\Phi'(0)$, so setting $g_{\mathrm{diff}}=0$ yields Eq.~\eqref{eq:direct-polarization-threshold}.

At an arbitrary equilibrium on an invariant axis,
\begin{equation}
  G'(h^*)
  =\beta\operatorname{sech}^2(\beta h^*)
  =\beta\left[1-(r^*)^2\right].
  \label{eq:app-tanh-derivative-equilibrium}
\end{equation}
The invariant-axis eigenvalues follow by substituting $F_x=-1$ and $F_r=1$ into Eqs.~\eqref{eq:consensus-eigenvalues} and \eqref{eq:balanced-eigenvalues}.

\subsection{Content-filtered updating}

For the content-filtered update law in Eq.~\eqref{eq:selective-update}, let
\begin{equation}
  u(x)=\tanh\!\left(\frac{k_Ax}{2}\right),
  \qquad
  u'(x)=\frac{k_A}{2}\left[1-u(x)^2\right].
  \label{eq:app-u-and-derivative}
\end{equation}
Differentiating $F_{\mathrm{sel}}$ gives
\begin{align}
  F_x(x,r)
  &=\frac{\gamma}{2}
  \left[
    u'(x)-1-r u(x)-x r u'(x)
  \right],
  \\
  F_r(x,r)
  &=\frac{\gamma}{2}
  \left[1-xu(x)\right].
  \label{eq:app-selective-F-derivatives}
\end{align}
At neutrality,
\begin{equation}
  A=\frac{\gamma}{2}
  \left(\frac{k_A}{2}-1\right),
  \qquad
  B=\frac{\gamma}{2}.
  \label{eq:app-selective-AB}
\end{equation}
Consequently,
\begin{align}
  g_{\mathrm{com}}
  &=\frac{\gamma}{2}
  \left\{
    \frac{k_A}{2}-1
    +\chi\left[\tau+(1-\tau)a\right]
  \right\},
  \\
  g_{\mathrm{diff}}
  &=\frac{\gamma}{2}
  \left\{
    \frac{k_A}{2}-1
    +\chi\left[\tau+(1-\tau)a\eta\right]
  \right\}.
  \label{eq:app-selective-neutral-rates}
\end{align}
The second line is the result quoted in Eq.~\eqref{eq:selective-difference-rate}.

\section{Numerical Identification and Continuation of Equilibria}
\label{app:numerical-continuation}

Asymmetric equilibria have no general scalar reduction and are therefore identified from the coupled equations
\begin{equation}
  \mathcal{M}(m,p)=0,
  \qquad
  \mathcal{P}(m,p)=0
\end{equation}
throughout the physical diamond $\mathcal{D}$. The numerical procedure used for the bifurcation diagrams consists of the following steps.

First, the stationary equations are solved from a collection of initial guesses covering $\mathcal{D}$. Roots outside the physical domain are discarded, and numerically coincident roots are merged using a prescribed tolerance. Second, the surviving equilibria are continued as the control parameter varies. Successive roots are matched using proximity in $(m,p)$, their symmetry class, and the signs of $m$ and $p$. Third, the complete Jacobian in Eq.~\eqref{eq:app-general-jacobian-expanded} is evaluated at every equilibrium. A point is classified as stable when both eigenvalues have negative real parts, as a saddle when $\det J<0$, and as unstable when $\det J>0$ and $\operatorname{tr}J>0$. Points with an eigenvalue sufficiently close to zero are retained as candidate bifurcation points and refined locally.

The symmetry-related copies of asymmetric equilibria are tracked explicitly. This procedure ensures that the continuation does not omit branches lying away from the invariant axes. In particular, it captures asymmetric saddles that organize basin boundaries even when they are not asymptotically stable.



\bibliography{pre_mean_field_references}

@article{Castellano2009,
  author  = {Castellano, Claudio and Fortunato, Santo and Loreto, Vittorio},
  title   = {Statistical physics of social dynamics},
  journal = {Rev. Mod. Phys.},
  volume  = {81},
  pages   = {591--646},
  year    = {2009},
  doi     = {10.1103/RevModPhys.81.591}
}

@article{Lorenz2007,
  author  = {Lorenz, Jan},
  title   = {Continuous opinion dynamics under bounded confidence: A survey},
  journal = {Int. J. Mod. Phys. C},
  volume  = {18},
  pages   = {1819--1838},
  year    = {2007},
  doi     = {10.1142/S0129183107011789}
}

@inproceedings{Deffuant2000,
  author    = {Deffuant, Guillaume and Neau, David and Amblard, Fr\'{e}d\'{e}ric and Weisbuch, G\'{e}rard},
  title     = {Mixing beliefs among interacting agents},
  booktitle = {Advances in Complex Systems},
  volume    = {3},
  pages     = {87--98},
  year      = {2000},
  doi       = {10.1142/S0219525900000078}
}

@article{HegselmannKrause2002,
  author  = {Hegselmann, Rainer and Krause, Ulrich},
  title   = {Opinion dynamics and bounded confidence: Models, analysis and simulation},
  journal = {J. Artif. Soc. Soc. Simul.},
  volume  = {5},
  number  = {3},
  pages   = {2},
  year    = {2002},
  url     = {https://www.jasss.org/5/3/2.html}
}

@article{SabinMillerAbrams2020,
  author  = {Sabin-Miller, David and Abrams, Daniel M.},
  title   = {When pull turns to shove: A continuous-time model for opinion dynamics},
  journal = {Phys. Rev. Research},
  volume  = {2},
  pages   = {043001},
  year    = {2020},
  doi     = {10.1103/PhysRevResearch.2.043001}
}

@inproceedings{Cornacchia2020,
  author    = {Cornacchia, Elisabetta and Singer, Neta and Abbe, Emmanuel},
  title     = {Polarization in attraction-repulsion models},
  booktitle = {2020 IEEE International Symposium on Information Theory},
  pages     = {2349--2354},
  year      = {2020},
  doi       = {10.1109/ISIT44484.2020.9174010}
}

@article{Baron2021,
  author  = {Baron, Joseph W.},
  title   = {Consensus, polarization, and coexistence in a continuous opinion dynamics model with quenched disorder},
  journal = {Phys. Rev. E},
  volume  = {104},
  pages   = {044309},
  year    = {2021},
  doi     = {10.1103/PhysRevE.104.044309}
}

@article{Gaisbauer2020,
  author  = {Gaisbauer, Felix and Olbrich, Eckehard and Banisch, Sven},
  title   = {Dynamics of opinion expression},
  journal = {Phys. Rev. E},
  volume  = {102},
  pages   = {042303},
  year    = {2020},
  doi     = {10.1103/PhysRevE.102.042303}
}

@article{Axelrod1997,
  author  = {Axelrod, Robert},
  title   = {The dissemination of culture: A model with local convergence and global polarization},
  journal = {J. Conflict Resolut.},
  volume  = {41},
  pages   = {203--226},
  year    = {1997},
  doi     = {10.1177/0022002797041002001}
}

@article{Sirbu2019,
  author  = {S\^{\i}rbu, Alina and Pedreschi, Dino and Giannotti, Fosca and Kert\'{e}sz, J\'{a}nos},
  title   = {Algorithmic bias amplifies opinion fragmentation and polarization: A bounded confidence model},
  journal = {PLoS ONE},
  volume  = {14},
  pages   = {e0213246},
  year    = {2019},
  doi     = {10.1371/journal.pone.0213246}
}

@article{Peralta2021,
  author  = {Peralta, Antonio F. and Neri, Matteo and Kert\'{e}sz, J\'{a}nos and I\~{n}iguez, Gerardo},
  title   = {Effect of algorithmic bias and network structure on coexistence, consensus, and polarization of opinions},
  journal = {Phys. Rev. E},
  volume  = {104},
  pages   = {044312},
  year    = {2021},
  doi     = {10.1103/PhysRevE.104.044312}
}

@techreport{Gans2024,
  author      = {Gans, Joshua S.},
  title       = {How Will Generative AI Impact Communication?},
  institution = {National Bureau of Economic Research},
  type        = {Working Paper},
  number      = {32690},
  year        = {2024},
  doi         = {10.3386/w32690}
}

@article{Matz2024,
  author  = {Matz, Sandra C. and Teeny, Jacob D. and Vaid, Sumer S. and Peters, H. and Harari, Gabriella M. and Cerf, Moran},
  title   = {The potential of generative AI for personalized persuasion at scale},
  journal = {Sci. Rep.},
  volume  = {14},
  pages   = {4692},
  year    = {2024},
  doi     = {10.1038/s41598-024-53755-0}
}

@article{Salvi2025,
  author  = {Salvi, Francesco and Ribeiro, Manoel Horta and Gallotti, Riccardo and West, Robert},
  title   = {On the conversational persuasiveness of GPT-4},
  journal = {Nat. Hum. Behav.},
  year    = {2025},
  doi     = {10.1038/s41562-025-02194-6}
}

@article{HruschkaAppel2026,
  author  = {Hruschka, Timon M. J. and Appel, Markus},
  title   = {Reducing political polarization through conversations with artificial intelligence},
  journal = {J. Comput.-Mediat. Commun.},
  volume  = {31},
  number  = {2},
  pages   = {zmag003},
  year    = {2026},
  doi     = {10.1093/jcmc/zmag003}
}

@book{BoydRicherson1985,
  author    = {Boyd, Robert and Richerson, Peter J.},
  title     = {Culture and the Evolutionary Process},
  publisher = {University of Chicago Press},
  address   = {Chicago},
  year      = {1985}
}

@article{HenrichMcElreath2003,
  author  = {Henrich, Joseph and McElreath, Richard},
  title   = {The evolution of cultural evolution},
  journal = {Evol. Anthropol.},
  volume  = {12},
  pages   = {123--135},
  year    = {2003},
  doi     = {10.1002/evan.10110}
}

@article{Stubbersfield2022,
  author  = {Stubbersfield, Joseph M.},
  title   = {Content biases in three phases of cultural transmission: A review},
  journal = {Cult. Evol.},
  volume  = {19},
  pages   = {41--60},
  year    = {2022},
  doi     = {10.1556/2055.2022.00024}
}

@article{Berl2021,
  author  = {Berl, Richard E. W. and Samarasinghe, Alarna N. and Roberts, Se\'{a}n G. and Jordan, Fiona M. and Gavin, Michael C.},
  title   = {Prestige and content biases together shape the cultural transmission of narratives},
  journal = {Evol. Hum. Sci.},
  volume  = {3},
  pages   = {e42},
  year    = {2021},
  doi     = {10.1017/ehs.2021.37}
}

@article{Buskell2019,
  author  = {Buskell, Andrew and Enquist, Magnus and Jansson, Fredrik},
  title   = {A systems approach to cultural evolution},
  journal = {Palgrave Commun.},
  volume  = {5},
  pages   = {131},
  year    = {2019},
  doi     = {10.1057/s41599-019-0343-5}
}

@article{Starnini2026,
  author  = {Starnini, Michele and Baumann, Fabian and Galla, Tobias and Garcia, David and I\~{n}iguez, Gerardo and Karsai, M\'{a}rton and Lorenz, Jan and Sznajd-Weron, Katarzyna},
  title   = {Opinion dynamics: Statistical physics and beyond},
  journal = {Rev. Mod. Phys.},
  volume  = {98},
  pages   = {035004},
  year    = {2026},
  doi     = {10.1103/j1zg-ddqv}
}

@article{Dandekar2013,
  author  = {Dandekar, Pranav and Goel, Ashish and Lee, David T.},
  title   = {Biased assimilation, homophily, and the dynamics of polarization},
  journal = {Proceedings of the National Academy of Sciences},
  volume  = {110},
  number  = {15},
  pages   = {5791--5796},
  year    = {2013},
  doi     = {10.1073/pnas.1217220110}
}

@article{Morin2016,
  author  = {Morin, Olivier},
  title   = {Reasons to be fussy about cultural evolution},
  journal = {Biology \& Philosophy},
  volume  = {31},
  number  = {3},
  pages   = {447--458},
  year    = {2016},
  doi     = {10.1007/s10539-016-9516-4}
}

@article{hancock2020aimediated,
    title = {{AI}-{Mediated} {Communication}: {Definition}, {Research} {Agenda}, and {Ethical} {Considerations}},
    volume = {25},
    shorttitle = {{AI}-{Mediated} {Communication}},
    url = {https://dx.doi.org/10.1093/jcmc/zmz022},
    doi = {10.1093/jcmc/zmz022},
    number = {1},
    urldate = {2026-03-31},
    journal = {Journal of Computer-Mediated Communication},
    publisher = {Oxford Academic},
    author = {Hancock, Jeffrey T. and Naaman, Mor and Levy, Karen},
    month = mar,
    year = {2020},
    pages = {89--100},
}

@article{brinkmann2023machine,
    title = {Machine culture},
    volume = {7},
    copyright = {2023 Springer Nature Limited},
    issn = {2397-3374},
    url = {https://www.nature.com/articles/s41562-023-01742-2},
    doi = {10.1038/s41562-023-01742-2},
    number = {11},
    urldate = {2025-03-22},
    journal = {Nature Human Behaviour},
    publisher = {Nature Publishing Group},
    author = {Brinkmann, Levin and Baumann, Fabian and Bonnefon, Jean-François and Derex, Maxime and Müller, Thomas F. and Nussberger, Anne-Marie and Czaplicka, Agnieszka and Acerbi, Alberto and Griffiths, Thomas L. and Henrich, Joseph and Leibo, Joel Z. and McElreath, Richard and Oudeyer, Pierre-Yves and Stray, Jonathan and Rahwan, Iyad},
    month = nov,
    year = {2023},
    pages = {1855--1868},
}

@article{flache2017models,
    title = {Models of social influence: {Towards} the next frontiers},
    volume = {20},
    issn = {14607425},
    url = {http://jasss.soc.surrey.ac.uk/20/4/2.html},
    doi = {10.18564/jasss.3521},
    number = {4},
    urldate = {2020-07-02},
    journal = {JASSS},
    publisher = {University of Surrey},
    author = {Flache, Andreas and Mäs, Michael and Feliciani, Thomas and Chattoe-Brown, Edmund and Deffuant, Guillaume and Huet, Sylvie and Lorenz, Jan},
    month = oct,
    year = {2017},
}

@article{mas2013differentiation,
    title = {Differentiation without {Distancing}. {Explaining} {Bi}-{Polarization} of {Opinions} without {Negative} {Influence}},
    volume = {8},
    issn = {1932-6203},
    url = {https://journals.plos.org/plosone/article?id=10.1371/journal.pone.0074516},
    doi = {10.1371/JOURNAL.PONE.0074516},
    number = {11},
    urldate = {2021-07-28},
    journal = {PLOS ONE},
    publisher = {Public Library of Science},
    author = {Mäs, Michael and Flache, Andreas},
    month = nov,
    year = {2013},
    pages = {e74516},
}

@article{jansson2026emergence,
    title = {The emergence of polarised groups through source filtering},
    volume = {13},
    doi = {10.1057/s41599-025-06419-x},
    journal = {Humanities and Social Sciences Communications},
    author = {Jansson, Fredrik and Hattiangadi, Anandi},
    year = {2026},
    pages = {112},
}

@article{claidiere2014how,
    title = {How {Darwinian} is cultural evolution?},
    volume = {369},
    url = {http://www.ncbi.nlm.nih.gov/pubmed/24686939},
    doi = {10.1098/rstb.2013.0368},
    number = {1642},
    urldate = {2018-11-26},
    journal = {Philosophical transactions of the Royal Society of London. Series B, Biological sciences},
    publisher = {The Royal Society},
    author = {Claidière, Nicolas and Scott-Phillips, Thomas C. and Sperber, Dan},
    month = may,
    year = {2014},
    pages = {20130368},
}

@article{jansson2021modelling,
    title = {Modelling {Cultural} {Systems} and {Selective} {Filters}},
    volume = {376},
    url = {https://doi.org/10.1098/rstb.2020.0045},
    doi = {10.1098/rstb.2020.0045},
    journal = {Philosophical Transactions of the Royal Society B: Biological Sciences},
    author = {Jansson, Fredrik and Aguilar, Elliot and Acerbi, Alberto and Enquist, Magnus},
    year = {2021},
    pages = {20200045},
}

@article{banisch2023biased,
    title = {Biased {Processing} and {Opinion} {Polarization}: {Experimental} {Refinement} of {Argument} {CommunicationTheory} in the {Context} of the {Energy} {Debate}},
    issn = {0049-1241},
    shorttitle = {Biased {Processing} and {Opinion} {Polarization}},
    url = {https://doi.org/10.1177/00491241231186658},
    doi = {10.1177/00491241231186658},
    urldate = {2024-06-03},
    journal = {Sociological Methods \& Research},
    publisher = {SAGE Publications Inc},
    author = {Banisch, Sven and Shamon, Hawal},
    month = jul,
    year = {2023},
    pages = {00491241231186658},
}

@misc{sharma2024generative,
    title = {Generative {Echo} {Chamber}? {Effects} of {LLM}-{Powered} {Search} {Systems} on {Diverse} {Information} {Seeking}},
    shorttitle = {Generative {Echo} {Chamber}?},
    url = {http://arxiv.org/abs/2402.05880},
    doi = {10.48550/arXiv.2402.05880},
    urldate = {2026-09-24},
    publisher = {arXiv},
    author = {Sharma, Nikhil and Liao, Q. Vera and Xiao, Ziang},
    month = feb,
    year = {2024},
    note = {arXiv:2402.05880 [cs.CL]},
}

@article{tessler2024ai,
    title = {{AI} can help humans find common ground in democratic deliberation},
    copyright = {Copyright © 2024 The Authors, some rights reserved; exclusive licensee American Association for the Advancement of Science. No claim to original U.S. Government Works},
    url = {https://www.science.org/doi/10.1126/science.adq2852},
    doi = {10.1126/science.adq2852},
    urldate = {2025-11-05},
    journal = {Science},
    publisher = {American Association for the Advancement of Science},
    author = {Tessler, Michael Henry and Bakker, Michiel A. and Jarrett, Daniel and Sheahan, Hannah and Chadwick, Martin J. and Koster, Raphael and Evans, Georgina and Campbell-Gillingham, Lucy and Collins, Tantum and Parkes, David C. and Botvinick, Matthew and Summerfield, Christopher},
    month = oct,
    year = {2024},
}

@article{jansson2026dynamics,
    title = {The dynamics of cultural systems},
    volume = {2601},
    url = {http://arxiv.org/abs/2601.00440},
    doi = {10.48550/arXiv.2601.00440},
    urldate = {2026-02-04},
    journal = {arXiv},
    publisher = {arXiv},
    author = {Jansson, Fredrik},
    month = jan,
    year = {2026},
    note = {arXiv:2601.00440 [physics]},
    pages = {00440},
}

@article{vandermaas2020polarization,
    title = {The polarization within and across individuals: the hierarchical {Ising} opinion model},
    volume = {8},
    issn = {2051-1329},
    shorttitle = {The polarization within and across individuals},
    url = {https://doi.org/10.1093/comnet/cnaa010},
    doi = {10.1093/comnet/cnaa010},
    number = {2},
    urldate = {2026-02-04},
    journal = {Journal of Complex Networks},
    author = {van der Maas, Han L J and Dalege, Jonas and Waldorp, Lourens},
    month = apr,
    year = {2020},
    pages = {cnaa010},
}

@article{hackenburg2024evaluating,
    title = {Evaluating the persuasive influence of political microtargeting with large language models},
    volume = {121},
    issn = {0027-8424},
    url = {https://pmc.ncbi.nlm.nih.gov/articles/PMC11181035/},
    doi = {10.1073/pnas.2403116121},
    number = {24},
    urldate = {2026-08-20},
    journal = {Proceedings of the National Academy of Sciences of the United States of America},
    author = {Hackenburg, Kobi and Margetts, Helen},
    year = {2024},
    pages = {e2403116121},
}

@article{bellina2023effect,
    title = {Effect of collaborative-filtering-based recommendation algorithms on opinion polarization},
    volume = {108},
    url = {https://link.aps.org/doi/10.1103/PhysRevE.108.054304},
    doi = {10.1103/PhysRevE.108.054304},
    number = {5},
    urldate = {2025-06-27},
    journal = {Physical Review E},
    publisher = {American Physical Society},
    author = {Bellina, Alessandro and Castellano, Claudio and Pineau, Paul and Iannelli, Giulio and De Marzo, Giordano},
    month = nov,
    year = {2023},
    pages = {054304},
}

@article{jansson2026modelling,
    title = {Modelling cultural evolution},
    volume = {2601},
    url = {http://arxiv.org/abs/2601.00433},
    doi = {10.48550/arXiv.2601.00433},
    urldate = {2026-02-04},
    journal = {arXiv},
    author = {Jansson, Fredrik},
    month = jan,
    year = {2026},
    note = {arXiv:2601.00433 [physics]},
    pages = {00433},
}

\end{document}